\documentclass[letterpaper,twocolumn,10pt]{article}
\usepackage{usenix2022_SOUPS}

\usepackage{tikz}
\usepackage{amsmath}
\usepackage{graphicx}
\usepackage{textcomp}
\usepackage{tabularx}
\newcolumntype{Y}{>{\centering\arraybackslash}X}

\newcolumntype{Z}{>{\hsize=1.3\hsize}X}
\newcolumntype{Q}{>{\hsize=.7\hsize}Y}
\newcolumntype{V}{>{\hsize=.15\hsize}X}
\usepackage{caption}
\usepackage{subcaption}
\usepackage{multirow}
\usepackage{multicol}
\usepackage{makecell}
\usepackage{xcolor}
\usepackage{xurl}
\usepackage{colortbl,hhline}
\usepackage{hyperref}
\usepackage{booktabs}

\usepackage{filecontents}

\begin{document}

\date{}

\title{\Large \bf Talking Abortion (Mis)information with ChatGPT on TikTok}

\def\plainauthor{Author name(s) for PDF metadata. Don't forget to anonymize for submission!}

\author{
{\rm Filipo Sharevski}\\
DePaul University
\and
{\rm Jennifer Vander Loop}\\
DePaul University
\and
{\rm Peter Jachim}\\
DePaul University
\and
{\rm Amy Devine}\\
DePaul University
\and
{\rm Emma Pieroni}\\
DePaul University
} 

\maketitle


\begin{abstract}
In this study, we tested users' perception of accuracy and engagement with TikTok videos in which ChatGPT responded to prompts about ``at-home'' abortion remedies. The chatbot's responses, though somewhat vague and confusing, nonetheless recommended consulting with health professionals before attempting an ``at-home'' abortion. We used ChatGPT to create two TikTok video variants - one where users can see ChatGPT explicitly typing back a response, and one where the text response is presented without any notion to the chatbot. We randomly exposed 100 participants to each variant and found that the group of participants unaware of ChatGPT's text synthetization was more inclined to believe the responses were misinformation. Under the same impression, TikTok itself attached misinformation warning labels (``\textit{Get the facts about abortion''}) to all videos after we collected our initial results. We then decided to test the videos again with another set of 50 participants and found that the labels did not affect the perceptions of abortion misinformation except in the case where ChatGPT explicitly responded to a prompt for a lyrical output. We also found that more than 60\% of the participants expressed negative or hesitant opinions about chatbots as sources of credible health information.

\end{abstract}

\section{Introduction}
Large language modeling of human-like dialog is a reality with several free ``chatbots'' available for people, researchers, coders, and test cheaters to use and experiment with \cite{Biderman}. Understandably, such a discursive sophistication and sensitivity attracts attention both in the usability and abusibility of these ``chatbots.''  Essentially Large-scale Language Models (LLM), chatbots are touted in their ability to fix bugs in programs \cite{Sobania}, give sound financial advice \cite{Wenzlaff}, and write impressive prose \cite{Looi}. But chatbots and LLMs could be triggered to produce malicious semantics for conveying offensive language \cite{Pan}, extract personally identifiable data \cite{Carlini}, and inject adversarial instructions to reproduce social biases and reinforce stereotypes \cite{Brown}. On the social engineering side, chatbots and LLMs have been successfully used to generate fake personas' resumes on popular social media sites \cite{Mink}, write phishing emails and social media posts \cite{Das, Seymour}, and produce a honeypot by issuing system commands \cite{McKee}.

As chatbots' objective is to produce persuasive language, they naturally make an attractive fit for an automated supply of misinformation and false narratives \cite{DiResta}. One could plausible generate a ``spin'' narrative against a person or topic of interest \cite{Bagdasaryan}, emulate the QAnon-style conspiratorial narratives \cite{Buchanan}, generate fake news \cite{Zellers}, and misuse rhetorical appeals to make a misinformation argument sound and credible \cite{Pauli}. Evidence suggests that people have difficulties distinguishing between chatbots and human-generated misinformation on political topics \cite{Kreps, Lin}, so it is question of time when the manual ``troll farms'' will be substituted with fully automated disinformation mills \cite{Huynh, Inie}. 

Chatbot-generated political misinformation certainly warrants close attention \cite{Hartmann}, but of equal, if not more pressing, importance is chatbot-generated health misinformation. Human-generated health misinformation, in the past, flooded social media \textit{en masse} with narratives that append a fear of either \textit{undesirable}, \textit{uncontrollable}, and \textit{unknown} health consequences. A prime example is the COVID-19 misinformation ``infodemic,'' seeded in part with previous disinformation about the Ebola and Zika viruses as well false information about the MMR vaccines \cite{Pullan, Wood-Zika, Pathak, Dixon}. Presently, human-generated health misinformation, appending the lack of \textit{desirable}, \textit{known} and \textit{accessible} health practices, is increasingly proliferated on social media in regards unproven treatments and ``at-home'' remedies such as alternative abortion treatments \cite{TikTok}. It is entirely plausible that in near future chatbots might replace humans in generating health misinformation, given the proliferation of chatbots for answering personal healthcare concerns \cite{Trec} and writing treatments on input diagnoses \cite{Jeblick}. 

Because misleading health information leads to vaccine hesitancy \cite{Loomba}, consumption of unproven remedies \cite{Southwell}, and attempts of unsafe procedures \cite{Kern}, it is an imperative to explore, first, how chatbots respond to health misinformation prompts, and second, how people respond to the chatbot answers to these prompts. We took upon this imperative and conducted a study on the topic of abortion misinformation with 150 participants. Inspired by the skyrocketing popularity of ChatGPT \cite{OpenAI} we queried this chatbot on four distinct prompts of ``at-home'' abortifacient herbs. We were particularly interested in abortion remedies because social media was abruptly flooded with questionable information about ``at-home'' attempts to induce miscarriage, following the US Supreme Court decision to strike down the legal right to abortion \cite{Totenberg}. We recorded the interaction with ChatGPT and posted the videos on TikTok, as previous research suggests that this platform in particular is a go-to place for social support exchange and questionable abortion information \cite{Barta, TikTok}.

We used two variants of the ChatGPT responses to each of our four ``at home'' abortion remedy prompts: (i) one where we recorded ChatGPT actually typing an response; and (ii) one in which the response of ChatGPT is superimposed verbatim over on a image as a static text (both variants are popular ways of formatting TikTok videos \cite{Southerton}). We randomly exposed 50 of the participants to four videos from the first variant and the other 50 to four videos from the second variant, asking their perception of accuracy of the statements in the videos. After we completed the main data collection of our study, we noticed that TikTok labeled all eight videos with a misinformation warning, urging the viewers to ``\textit{get the facts about abortion}.'' Given one in three users ignore the abortion misinformation labels on TikTok \cite{TikTok}, we followed up with a second round of data collection with 50 new participants now with an explicit directive by TikTok that the videos might not be factually correct in regards ``at-home'' abortion remedies. 

We prompted ChatGPT to respond about abortifacient herbs, to say something about herbs used for abortion, to say the facts about using herbs for abortion, and to write lyrics about these herbs (the prompts and responses are  given in the \hyperref[sec:survey]{Appendix}). The ChatGPT's responses, expectedly, did not contain explicit misinformation though they hinted that herbs could be used for inducing abortions. Each response presented a rather encyclopedic and general answer to the prompts and at urged the users to consult with a medical professional before using any of the abortifacient herbs for inducing abortion on their own. The lyrical answer extolled the virtue of a particular abortifacient herb, pointing its use for menstrual cramps, but did not mention its explicit use for abortion (the menstrual cramps benefit is the main justification for recommending the use of this herb as an abortifacient ``at-home'' remedy \cite{Pennyroyal-oil}). 

We found that participants who weren't aware that the videos were created by ChatGPT were more inclined to perceive the textual response as misinformation for all for prompts. The participants perceiving the ChatGPT responses as factually inaccurate pointed out that the chatbot omits references on toxicity and dosage therefore spreading ``\textit{misinformation by omission}.'' The impression of misinformation was also justified in the formatting of the responses that, in the words of the participants, ''\textit{were biased towards selling you products, not informing you about it}.'' The misinformation label, perhaps overcautious and algorithmically assigned by TikTok, made little change in the perception of accuracy except for the lyrical prompt where it nudged a higher number of participants to deem the videos as misinformation.

To report the findings from our study, we reviewed the prior work on misinformation relative to LLMs, and generative health (mis)information in Section \ref{sec:chatbots-misinfo}. Section \ref{sec:abortion-misinfo} provides the broader context of health and abortion misinformation narratives on social media. Section \ref{sec:study} covers the methodological details of our study. Section \ref{sec:results} elaborates how participants assessed and engaged with abortion misinformation, and Section \ref{sec:results-reception} the receptivity of our participants to chatbot generated abortion responses, respectively. We draw on our findings in Section \ref{sec:discussion} to discuss the implications for ``neural'' abortion (mis)information as well as the relative content moderation on social media. Finally, Section \ref{sec:conclusion} concludes the paper.

\section{Synthetic Misinformation} \label{sec:chatbots-misinfo}
\subsection{LLMs and Neural Fake News}
The synthetic or ``neural'' fake news, early on, were seen a serious threat to the constrictive discourse online \cite{Solaiman, Weidinger}. Highlighting the plausibility of this threat, Zellers et al. showed that a synthetic propaganda is perceived as more believable than human-generated propaganda among human readers \cite{Zellers}. Similarly, Newhouse et al. demonstrated that LLMs are able to generate text with an ideological consistency from any extremist view used an input \cite{Newhouse}. The ability to generate such a trolling content \textit{en masse}, thus, is naturally appealing for state sponsored and any other malicious trolls that so far had to manually manufacture fake news, rumors, and conspiracy theories around polarising issues \cite{TrollHunter}.

Goldstein et al. argue that ``neural'' fake news will (a) drive down the costs of operating any troll farms, allowing for new ones to quickly appear; (b) enable fast scaling and cross-platform testing; and (c) improve the linguistic or cultural inconsistencies innate to the human-generated  misinformation \cite{Goldstein}. This, in turn, will lower the existing barriers and complicate detection of state sponsored trolling, given that they can fuse the synthetic misinformation with their proven abilities to pose as authentic, culturally competent personas (e.g. the so-called ``Jenna Abrams'' accounts \cite{Xia}) or vocal supporters of hashtag activism movements (e.g. \@BlackToLive in \#BlackLivesMatter \cite{Stewart}). 

An early proof of this concerning scenario is the LLM trained on \texttt{4chan} -- dubbed GPT-4chan -- to produce offensive and hateful yet very believable synthetic trolling narratives \cite{Murphy}. Compared to the past influence operations \cite{DiResta}, the GPT-4chan troling is not just comparably cruel in sentiment, but dangerously more powerful in volume output \cite{Lu}. Measuring the toxicity of chatbots' content,  Si et al. found that popular chatbots like BlenderBot and TwitterBot \cite{Parlai} are not just prone to providing toxic responses -- offensive language that involves hateful or violent content -- when fed toxic queries (e.g. commentary from 4chan and Reddit), but non-toxic prompts can trigger such responses too \cite{Si}. Such synthetic responses, Schuster et al. show, complicate the detection of online trolling and fake news as they are harder to detect due to the excellent stylometric obfuscation \cite{Schuster}. 

\subsection{Chatbots and Health (Mis)information} 
Chatbots, facilitated by language modeling, became the go-to way of addressing accessibility issues with traditional in-person healthcare with the proliferation of smartphones and broadband internet access \cite{Bates, Abedin}. Healthcare conversational agents provide answers relative to mental health \cite{Easton}, cancer \cite{Siglen}, viruses \cite{Miner}, substance abuse \cite{Laranjo}, and even help to dispel COVID-19 misinformation \cite{Xiao}. A convincing evidence on how the new and powerful chatbots like ChatGPT perform in regards providing general medical advice is still absent, though ChatGPT already showed a good performance when taking the medical licensing exam in the US \cite{Kung, Gilson}. 

The generative capabilities of ChatGPT were sufficient in writing factually correct radiology reports \cite{Jeblick}, correctly answer cirrhosis and hepatocellular carcinoma questions \cite{Yeo}, provide accurate pediatric diagnosis \cite{Benoit}, and convincingly arguing for using AI-based chatbots in providing care to patients \cite{D'Amico, King, Liebrenz}. Though promising aiding healthcare decision-making, the latest generation of chatbots are yet to be evaluated relative to generating falsehoods, misleading claims, and speculative treatment. Such capabilities for synthetic health misinformation are deliberately restricted in the design and development of these chatbots  \cite{Solaiman}, however, they could be skilfully triggered to ``hallucinate'' i.e. fabricate a credible but incorrect medical advice \cite{Rao}. Additionally, the ChatGPT and the likes are currently only evaluated by highly experienced healthcare and LLM researchers \cite{Lin}, without any evidence on how ordinary users and patients perceive, incorporate, and act upon the synthesized medical advice.

\section{Abortion Misinformation} \label{sec:abortion-misinfo}
So far, ChatGPT has not been formally tested in providing abortion information as a medical advice, though the chatbot could persuasively argue for legalizing abortion in US \cite{Hartmann}. This particular capability of ChatGPT draws relevance, perhaps in part, to the polarized abortion discourse online abruptly amplified in the immediate aftermath of the US Supreme Court decision to strike down the constitutional right for abortion \cite{Totenberg}. The inability to obtain a legal abortion turned people to search engines and social media to learn how to manage their reproductive decisions and perform safe abortions \cite{Sherman}. Unfortunately, not all information aligned with the National Library of Medicine’s description of abortion and recommendations for safe practices \cite{Ajmal}. 

\subsection{Post \textit{Roe vs Wade} ``At-Home'' Remedies}

Unlike the vaccination misinformation narratives, driven by a fear mongering conspiratorial narratives targeting vaccine \textit{hesitancy} \cite{Linden}, the abortion misinformation is driven by a reproductive \textit{resolution} to try untested ``at-home'' abortion remedies \cite{Spencer}. In the post Roe vs Wade discourse, for example, many questionable ``at-home'' practices including pills, oils, and herbs for inducing abortion flooded social media, both as claims and as an advertisements in users' feeds \cite{Sherman}. As prior evidence shows that 70.1\% of women obtain information regarding abortion from the Internet \cite{Littman}, it is likely that these misleading claims will not just show in many users' feeds, but that some users will pursue the relative treatments. Reports, anecdotally, already show that women have been admitted in emergency rooms seeking critical lifesaving treatment following failed ``at-home'' attempts to induce abortion \cite{RubinR}. 

Abortion misinformation online, literature shows, takes many forms and users generally have difficulties discerning inaccuracies in the related alternative treatments \cite{Patev}. The inability to spot falsehoods relative to the safety, infertility, mental health risk, and legality of abortion \cite{Bessett} is a cause for serious concern as reports indicated that abortion misinformation specifically related to an ``abortion reversal pill'' increased on Facebook from 20 interactions on June 23 to 3,500 interactions on June 24 2022, the day after the Supreme Court decision to overturn \textit{Roe v. Wade}. \cite{Kern}. The momentum of concern is even more evident as the Spanish-language abortion misinformation was deliberately designed to galvanize voters in Latino communities across the US, following the Supreme Court ruling \cite{Godoy}.

Abortifacient herbs -- purportedly providing the ability to induce a spontaneous miscarriage -- form the majority of post-\textit{Roe v Wade} misinformation \cite{Swenson}. The toxicity of abortifacient herbs like  has been widely studied, alas, without of an explicit effect in inducing ``at-home'' abortions. \cite{Johnson-Arbor}. Most existing studies were done in countries other than the US, where abortion was not legal until recently. Abortion did not become legal in Uruguay until 2012 \cite{Makleff}, for example, and a 2003 study found that the Montevideo Poison Centre had 86 cases of ingestion of herbal infusions with abortive intent from 1986 to 1999 \cite{Ciganda}. In the United States, misinformation surrounding ``herbal abortions''  has increased dramatically on social media after the legal abortion was overturned, especially in viral videos on TikTok \cite{Case}. 

\subsection{Social Media Handling}
Platforms used diverse strategies to mitigate abortion misinformation: YouTube added ``context labels'' to such abortion content \cite{YouTube}, Twitter decided to promote authoritative abortion information in its Twitter Moments and Events \cite{Kern}, and Meta purportedly blocked questionable abortion treatment advertisements \cite{Meta}. TikTok also stated it removed and labeled videos with abortion misinformation \cite{Keenan}, but many of the questionable home practices aimed to ``cause a miscarriage'' still appeared in users' personal streams \cite{newsguard}. Debunking of abortion misinformation on TikTok followed up \cite{Spencer}, but the slow-in-nature checking and verifying of health-related facts was no match for the rapid spread of videos recommending dangerous abortion remedies.

TikTok -- deemed the ``New Google'' for Gen-Z \cite{Gupta} -- draws special attention relative to abortion misinformation, pressing reproductive decisions are particularly interesting to the majority of users on this platform. TikTok's status as a platform for social support exchange \cite{Barta} further exacerbates the immediate danger of abortion remedies as supportive communication adds to ``stickiness'' and internalization of such content among adolescents and young adults \cite{Duggan}. Studies focused on abortion misinformation on TikTok already show that this danger is real as roughly 30\% of the users believed in ``at-home'' remedies' safety and efficacy, despite those being already scientifically debunked \cite{TikTok}. Worse, the explicitly debunking label attached to a misleading abortion video by TikTok about the harms of ``at-home'' did not help a third of the users in the study to dismiss a video about self-administering abortion as misinformation.

\section{Study Design} \label{sec:study}
\subsection{Research Questions}
As fully LLM powered chatbots like ChatGPT will inevitably appear in many health-related conversations and online searches \cite{Goldstein}, it is an imperative to learn how the \textit{users} perceive, assess and engage with synthesized responses. We took upon this imperative in the context of abortion ``at-home'' remedies because internalizing the related misinformation has immediate dangers to the well-being of the users, whom abruptly were restricted access to reproductive healthcare by the US Supreme Court decision. Therefore, we set to answer the following research questions in our study: 

\vspace{0.5em}
\begin{itemize}
\itemsep 0.5em
    \item \textbf{RQ1a:}\ \textit{Assessment}: How do TikTok users assess ChatGPT responses relative to ``at-home'' abortion remedies prompts in videos explicitly showing the interaction with the chatbot?

    \item \textbf{RQ1b:}\ \textit{Assessment}: How do TikTok users assess ChatGPT responses relative to ``at-home'' abortion remedies prompts in videos showing only the textual response? 
    
     \item \textbf{RQ2:}\ \textit{Engagement}: What strategies users employ in assessing and responding to ``at-home'' abortion videos created with ChatGPT on TikTok?

    \item \textbf{RQ3:}\ \textit{Reception}: What is users' general reception of information generated by ChatGPT and language models?
    
\end{itemize}

\subsection{Sample}
We obtained an IRB approval for fielding an anonymous, exploratory study where ordinary users directly interacted on TikTok with short videos either showing an interaction with ChatGPT or the chatbot's response to ``at-home'' abortion remedies' prompts. After the interaction, users answered series of questions, given in the \hyperref[sec:survey]{Appendix}, relative to the accuracy of the information in the videos, experience with chatbots, and their engagement strategies with misinformation. We sampled TikTok users ages 18 and above in the United States and used Prolific for recruitment. Our participants were allowed to skip any question they were uncomfortable answering, taking around 15 minutes to complete interact with a random selection of the videos and complete the survey. Participants were offered a compensation rate of \$3 each. Following a preliminary power analysis and a data consolidation, we ended with an initial sample of 100 participants. The sample's demographic distribution is given in Table \ref{tab:demographics}.

\begin{table}[htbp]
\renewcommand{\arraystretch}{1.3}
\footnotesize
\caption{Initial Sample Demographic Distribution $N=100$}
\label{tab:demographics}
\centering
\aboverulesep=0ex 
\belowrulesep=0ex 
\begin{tabularx}{0.8\linewidth}{|Y|}
\Xhline{3\arrayrulewidth}
\toprule
 \textbf{Gender} \\\Xhline{3\arrayrulewidth}
\midrule
\footnotesize
\vspace{0.2em}
    \hfill \makecell{\textbf{Female} \\ 70 (70\%)} 
    \hfill \makecell{\textbf{Male} \\ 24 (24\%)} 
    \hfill \makecell{\textbf{Non-cisgender} \\ 6 (6\%)} \hfill\null
\vspace{0.2em}
\\\Xhline{3\arrayrulewidth}
\midrule
 \textbf{Age} \\\Xhline{3\arrayrulewidth}
\midrule
\footnotesize
\vspace{0.2em}
\hfill \makecell{\textbf{[18-20]}\\ 9 (9\%)}
    \hfill \makecell{\textbf{[21-30]} \\ 53 (53\%)} 
    \hfill \makecell{\textbf{[31-40]}\\ 23 (23\%)} 
    \hfill \makecell{\textbf{[41-50]} \\ 5 (5\%)}
    \hfill \makecell{\textbf{[51-60]} \\ 6 (6\%)}
    \hfill \makecell{\textbf{[61+]} \\ 4 (4\%)}
\vspace{0.2em}
\\\Xhline{3\arrayrulewidth}
\midrule
 \textbf{Political leanings} \\\Xhline{3\arrayrulewidth}
\midrule
\footnotesize
\vspace{0.2em}
    \hfill \makecell{\textbf{Left} \\ 58 (58\%)} 
    \hfill \makecell{\textbf{Moderate} \\ 25 (25\%)} 
    \hfill \makecell{\textbf{Right} \\ 12 (12\%)} 
    \hfill \makecell{\textbf{Apolitical} \\ 5 (5\%)} \hfill\null 
\vspace{0.2em}
\\\Xhline{3\arrayrulewidth}
\midrule
 \textbf{Highest Level of Education Completed} \\\Xhline{3\arrayrulewidth}
\midrule
\footnotesize
\vspace{0.2em}
    \hfill \makecell{\textbf{High school} \\ 19 (19\%)} 
    \hfill \makecell{\textbf{College} \\ 71 (71\%)} 
    \hfill \makecell{\textbf{Graduate} \\ 10 (10\%)} \hfill\null 
\vspace{0.2em}
\\\Xhline{3\arrayrulewidth}
\bottomrule
\end{tabularx}
\end{table}

After we completed the main data collection of our study, we noticed that TikTok labeled all eight videos with a misinformation warning to ``\textit{get the facts about abortion}.'' According to TikTok's safety policy, a proactive detection program substantiated through fact-checking flags new and evolving claims in regarding the most popular topics of misinformation -- vaccines, abortion, and voting \cite{Keenan}. As the abortion misinformation labels on TikTok haven't yielded the anticipated dispelling effect with 30\% of users in a previous test \cite{TikTok}, we initiated a second round of data collection with 50 new participants to collect their impressions of abortion advice given by ChatGPT, now with an explicit directive by TikTok that this advice might not be factually correct in regards ``at-home'' remedies. Using the same data collection setup, we recruited an additional sample of 50 TikTok users, with a demographic distribution is given in Table \ref{tab:demographics-addition}.

\begin{table}[htbp]
\renewcommand{\arraystretch}{1.3}
\footnotesize
\caption{Additional Sample Demographic Distribution $N=50$}
\label{tab:demographics-addition}
\centering
\aboverulesep=0ex 
\belowrulesep=0ex 
\begin{tabularx}{0.8\linewidth}{|Y|}
\Xhline{3\arrayrulewidth}
\toprule
 \textbf{Gender} \\\Xhline{3\arrayrulewidth}
\midrule
\footnotesize
\vspace{0.2em}
    \hfill \makecell{\textbf{Female} \\ 35 (70\%)} 
    \hfill \makecell{\textbf{Male} \\ 13 (26\%)} 
    \hfill \makecell{\textbf{Non-cisgender} \\ 2 (4\%)} \hfill\null
\vspace{0.2em}
\\\Xhline{3\arrayrulewidth}
\midrule
 \textbf{Age} \\\Xhline{3\arrayrulewidth}
\midrule
\footnotesize
\vspace{0.2em}
\hfill \makecell{\textbf{[18-20]}\\ 8 (16\%)}
    \hfill \makecell{\textbf{[21-30]} \\ 25 (50\%)} 
    \hfill \makecell{\textbf{[31-40]}\\ 11 (22\%)} 
    \hfill \makecell{\textbf{[41-50]} \\ 3 (6\%)}
    \hfill \makecell{\textbf{[51-60]} \\ 2 (4\%)}
    \hfill \makecell{\textbf{[61+]} \\ 1 (2\%)}
\vspace{0.2em}
\\\Xhline{3\arrayrulewidth}
\midrule
 \textbf{Political leanings} \\\Xhline{3\arrayrulewidth}
\midrule
\footnotesize
\vspace{0.2em}
    \hfill \makecell{\textbf{Left} \\ 30 (60\%)} 
    \hfill \makecell{\textbf{Moderate} \\ 12 (24\%)} 
    \hfill \makecell{\textbf{Right} \\ 5 (10\%)} 
    \hfill \makecell{\textbf{Apolitical} \\ 3 (4\%)} \hfill\null 
\vspace{0.2em}
\\\Xhline{3\arrayrulewidth}
\midrule
 \textbf{Highest Level of Education Completed} \\\Xhline{3\arrayrulewidth}
\midrule
\footnotesize
\vspace{0.2em}
    \hfill \makecell{\textbf{High school} \\ 5 (10\%)} 
    \hfill \makecell{\textbf{College} \\ 39 (78\%)} 
    \hfill \makecell{\textbf{Graduate} \\ 6 (12\%)} \hfill\null 
\vspace{0.2em}
\\\Xhline{3\arrayrulewidth}
\bottomrule
\end{tabularx}
\end{table}

\subsection{Method and Analysis}
Participants were provided an open ended qualitative survey through Qualtrics that provided a list of questions and a predetermined set of TikTok videos. We used prompts in ChatGPT to generate text regarding herbal abortions in general and pennyroyal in particular. Various herbs are offered as ``at-home'' abortion remedies with claims that they have ``abortifacient'' effect i.e. induce miscarriage, such as blue/black cohosh, eastern daisy fleebane, mugworth, parsley, pennyroyal and rue \cite{Johnson-Arbor}. Three independent members of the research team spent an extensive time prompting ChatGPT with each of the abovementioned abortifacient herbs, but only pennyroyal was the one that came up each time regardless of what prompt text, device (e.g. computer, smartphone), browser, time of the day, or location was used.

\begin{figure*}[!h]
  \centering
\begin{subfigure}[t]{.3\linewidth}
    \centering\includegraphics[width=.6\linewidth]{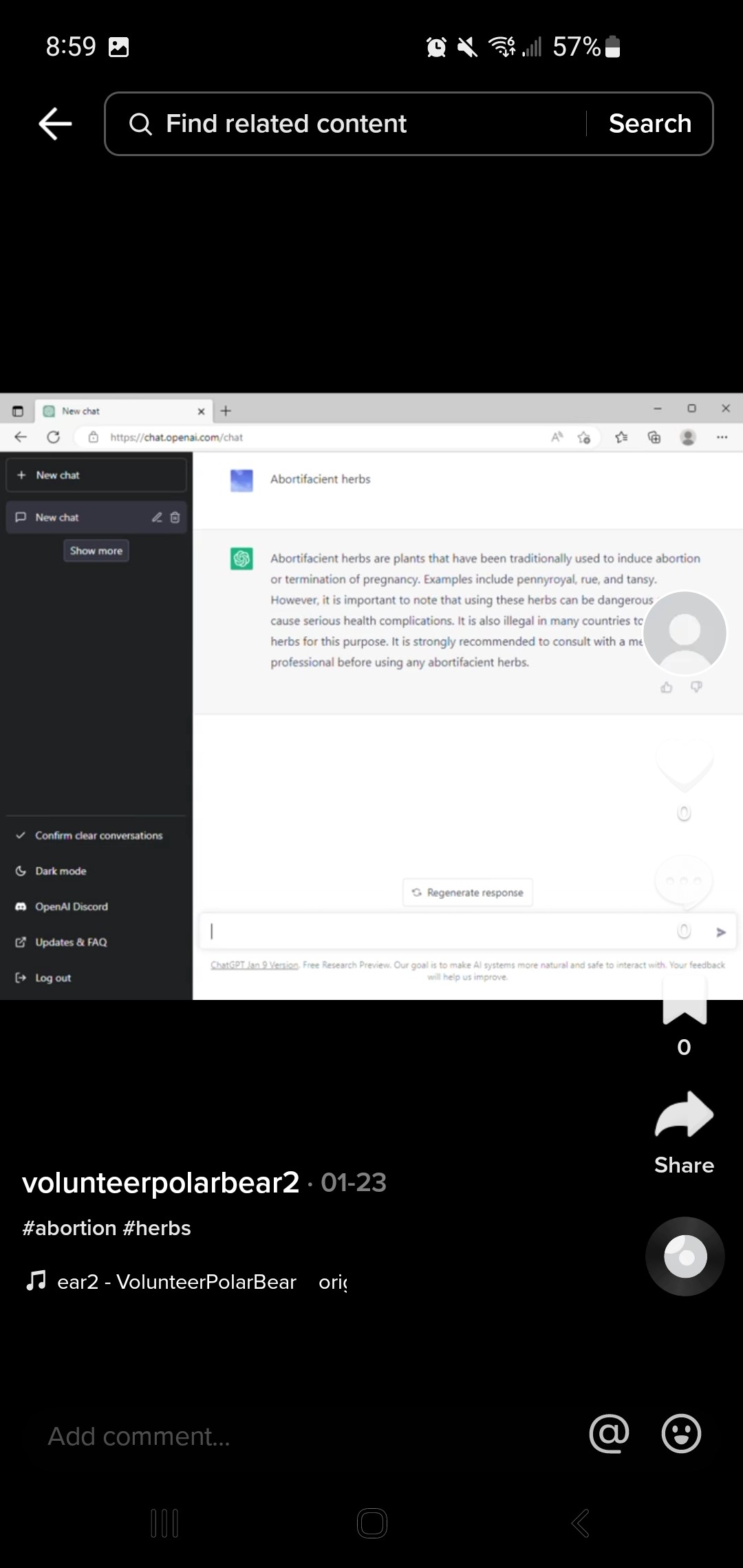}
    \caption{Explicit ChatGPT video}
    \label{fig:explicit}
  \end{subfigure}
  \begin{subfigure}[t]{.3\linewidth}
    \centering\includegraphics[width=.6\linewidth]{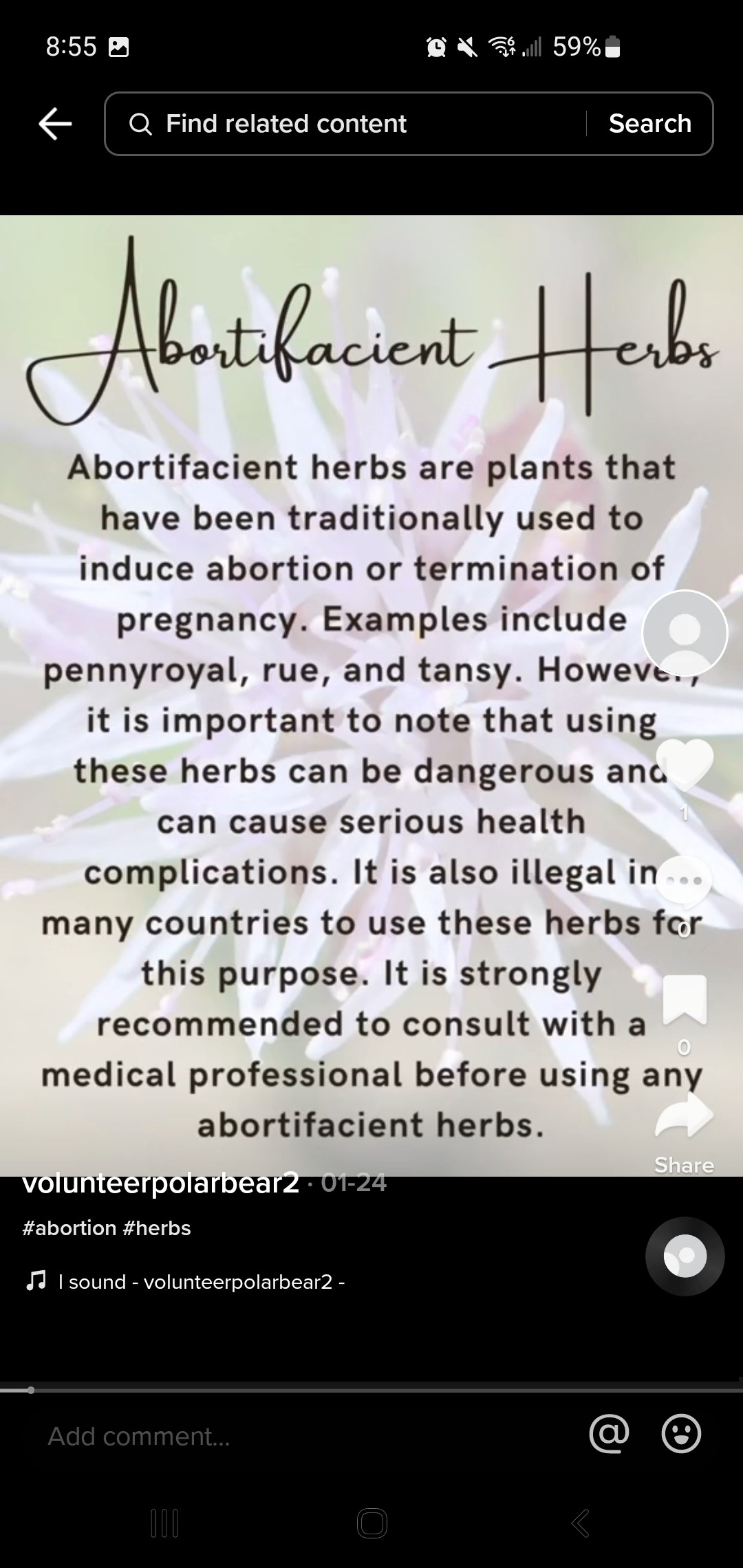}
    \caption{Implicit ChatGPT video}
    \label{fig:implicit}
  \end{subfigure}
  \begin{subfigure}[t]{.3\linewidth}
    \centering\includegraphics[width=.6\linewidth]{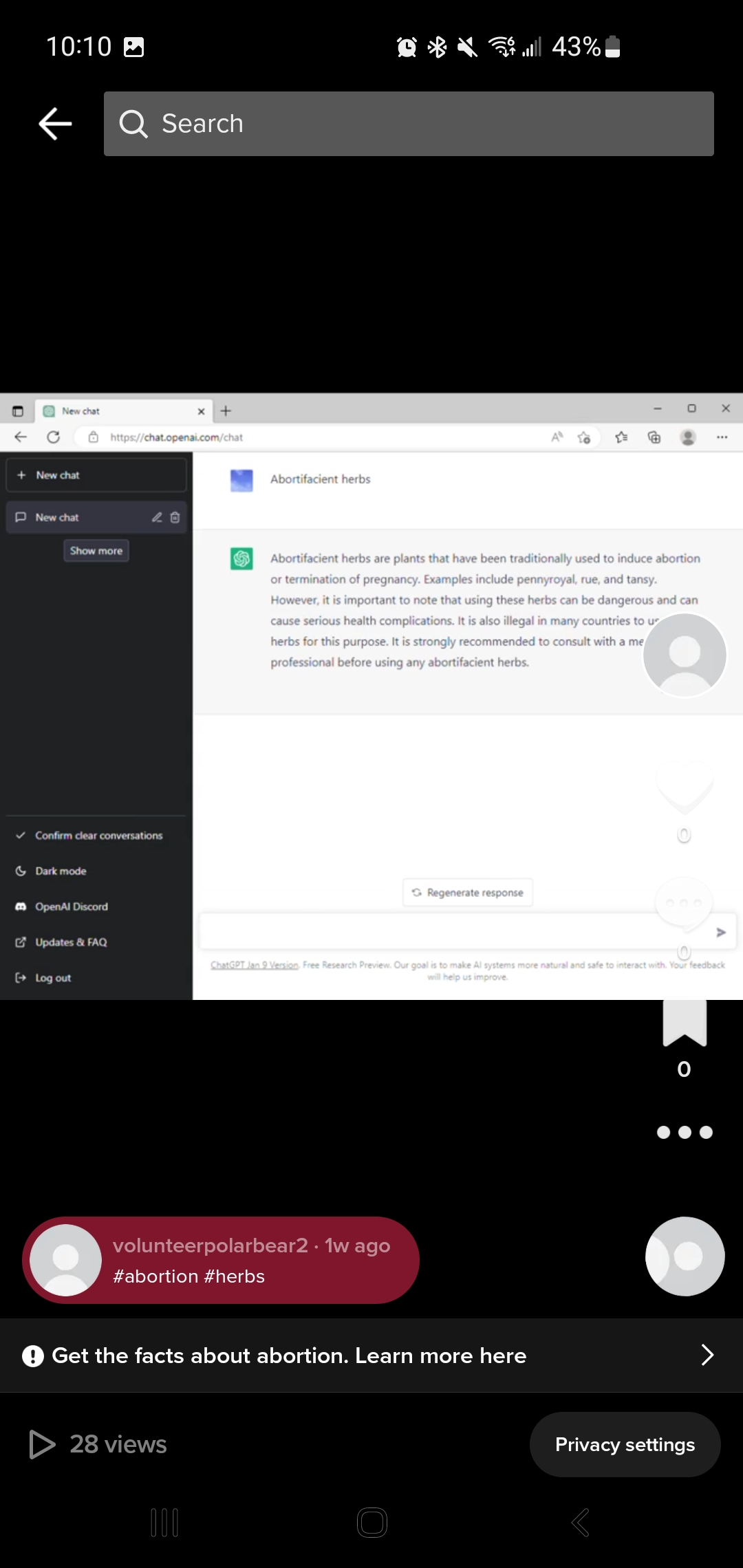}
    \caption{Misinformation Labeled video}
    \label{fig:labeled}
  \end{subfigure}
  \caption{TikTok videos used in the study: Screenshots} 
\end{figure*}

The prompts and ChatGPT responses we selected to use in our study are provided in the \hyperref[sec:ChatGPT]{Appendix}. We then used the conversational results to create a total of eight short TikTok videos. Four of the videos showed a prompt being entered into ChatGPT and the response being generated on the website as depicted in the screenshot in Figure \ref{fig:explicit}. The other four show only the response that was generated through ChatGPT overlaying an image of the herb pennyroyal and did not indicate that the text had been created using a chatbot, as depicted in Figure \ref{fig:implicit}. The videos were posted from an innocuous account without an avatar or any history of posting other videos to avoid any confounding effects. Participants were randomly assigned into two groups of 50 each. In the first group, participants were each shown four randomly selected videos explicitly showing the conversational interaction with ChatGPT. The second group of participants were also shown four randomly selected videos that contained only the text that had been generated by ChatGPT. We randomly selected the exposure to the videos in both groups to avoid habituation and question order bias in evaluation of the content.

After the initial survey concluded, TikTok added misinformation labels on the videos in the mobile application version of TikTok only, as shown in Figure \ref{fig:labeled}, linked to the National Library of Medicine's MedlinePlus webpage on abortion \cite{Mediline}. For the additional data collection we stipulated a use the mobile application which showed them the TikTok misinformation label. We then randomly divided the participants into two groups of 25. The first group viewed the set of four videos showing ChatGPT generating the text and the second group viewed the videos that only showed the overlaying text.

Two independent researchers coded and analyzed the results using the codebook in \cite{TikTok} and achieving a strong level of inter-coder agreement (Cohen's $\kappa = .87$). We utilized a thematic analysis methodology to identify the assessment themes most saliently emerging from the responses in our sample. The themes were summarized to describe the subjective  assessment of the ChatGPT's responses, the relative engagement actions taken (e.g. scroll past, fact-check, block, report, reply, like), and the general experience with chatbots. In reporting the results, we utilized as much as possible verbatim quotation of participants' answers, emphasized in ``\textit{italics}'' and with a reference to the participant as either \textbf{PXYZ\#} or [\textbf{PXYZ\#}], where \textbf{P} denotes \textbf{participant}, \textbf{X} denotes the \textbf{number} of the participant in the sample (ordered by the time of participation), \textbf{Y} denotes their \textbf{gender} identity (\textbf{F} - female, \textbf{M} - male, \textbf{NC} - non-cisgender), \textbf{Z} denotes their \textbf{political} identity (\textbf{L} - left-leaning, \textbf{M} - moderate, \textbf{R} - right-leaning; \textbf{A} - apolitical), and \textbf{\#} denotes the upper bound of their \textbf{age bracket}. For example, \textbf{P16FL30} refers to \textbf{participant 16}, \textbf{female}, \textbf{left-leaning}, \textbf{age bracket} \textbf{[21-30]}.

\section{Results: Assessment and Engagement} \label{sec:results}
\subsection{Abortifacient herbs}
The first set of TikTok videos, shown in Appendix \ref{sec:video1}, were created using the prompt ``\textit{Abortifacient herbs}.'' As shown in Table \ref{tab:video1-assessment}, 25 participants were randomly selected to view the explicit ChatGPT video and 26 to view the implicit ChatGPT video. Thirteen participants from the explicit ChatGPT group (52\%) stated they ``\textit{don't know for sure if the video contains any misinformation because these herbs could make an abortion happen}'' [\textbf{P24FL30}], but  they ``\textit{try not to take anything as fact on TikTok}'' [\textbf{P31FL40}]. Seventeen of participants in the implicit group (65.38\%) were  ``\textit{not sure if the post is misinformation but [they]] will give them credit that they push the reader to go to their healthcare provider}'' [\textbf{P82FL30}]. Nine participants from the explicit ChatGPT group (36\%) stated that ``\textit{it doesn't look like there is misinformation about it}'' [\textbf{P2FL40}] added they felt this way ``\textit{because of the examples provided}'' [\textbf{P7FL20}]. Three (12\%) participants from the explicit ChatGPT group  the video  ``\textit{contains misinformation, giving medical advice about abortion, is very dangerous}'' [\textbf{P45FM40}] and four (of participants from the implicit ChatGPT group 15.38\%)  state they ``\textit{believe it has misinformation since it is from a random account and not a doctor/scientist who would know this information}'' [\textbf{P83MM30}].

\label{tab:video1-assessment}
\begin{table}[htbp]
\renewcommand{\arraystretch}{1.3}
\footnotesize
\caption{Is Video \#1 Misinformation?}
\label{tab:video1-assessment}
\centering
\aboverulesep=0ex 
   \belowrulesep=0ex 
\begin{tabularx}{\linewidth}{|Y|}
\Xhline{3\arrayrulewidth}
\toprule
\textbf{Explicit ChatGPT Group} [\textbf{viewed}: 25 participants]
\\\Xhline{3\arrayrulewidth}
\midrule
\footnotesize
\vspace{0.2em}
    \hfill \makecell{\textbf{Yes} \\ 3 (12\%)} 
    \hfill \makecell{\textbf{No} \\ 9 (36\%)} 
    \hfill \makecell{\textbf{Unsure} \\ 13 (52\%)} \hfill\null
\vspace{0.2em}
\\\Xhline{3\arrayrulewidth}
\midrule
 \textbf{Implicit ChatGPT Group} [\textbf{viewed}: 26 participants] \\\Xhline{3\arrayrulewidth}
\midrule
\footnotesize
\vspace{0.2em}
    \hfill \makecell{\textbf{Yes} \\ 4 (15.38\%)} 
    \hfill \makecell{\textbf{No} \\ 6 (23.08\%)} 
    \hfill \makecell{\textbf{Unsure} \\ 16 (61.54\%)} \hfill\null
\vspace{0.2em}
\\\Xhline{3\arrayrulewidth}
\bottomrule
\end{tabularx}
\end{table}

While the majority of the explicit ChatGPT group participants were inclined to scroll past the first video, as shown in Table \ref{tab:video1-action}, 16\% or four of them were keen on fact-checking ChatGPT's response, saying they ``\textit{would take the time to research the info [they] found before taking any action}'' [\textbf{P3FL30}]. Two other participants were keen on replying and reporting the explicit video because ``\textit{giving medical advice about abortion on TikTok is very dangerous}'' [\textbf{P45FM40}]. Participants in the implicit ChatGPT group were more likely to take action on the first video as seven of them (26.92\%) pointed they would perform an ``\textit{educated research and if severe risks were at stake I may have been inclined to comment or even possibly provide a reputable link counter this information}'' [\textbf{P92FL40}]. Two participants said they would block the video account and another two were keen to  ``\textit{flag it for dangerous information}'' [\textbf{P90FM20}]. One participant stated they would comment because ``\textit{this seems irrational or misinformation and it should not be spread on TikTok giving girls/women ideas of a dangerous act}'' [\textbf{P70FM40}] and one liked the video because ``\textit{it is addressing what the herbs are while also educating people about how harmful they are}'' [\textbf{P62FL30}].

\begin{table}[htbp]
\renewcommand{\arraystretch}{1.3}
\footnotesize
\caption{What action would you take on Video \#1?}
\label{tab:video1-action}
\centering
 \aboverulesep=0ex 
   \belowrulesep=0ex 
\begin{tabularx}{\linewidth}{|Y|}
\Xhline{3\arrayrulewidth}
\toprule
 \textbf{Explicit ChatGPT Group} [\textbf{viewed}: 25 participants] \\\Xhline{3\arrayrulewidth}
\midrule
\footnotesize
\vspace{0.2em}
    \hfill \makecell{\textbf{Scroll Past} \\ 19 (76\%)} 
    \hfill \makecell{\textbf{Fact-check} \\ 4 (16\%)} 
    \hfill \makecell{\textbf{Block} \\ 0 (0\%)} 
    \hfill \makecell{\textbf{Report} \\ 1 (4\%)}
    \hfill \makecell{\textbf{Reply} \\ 1 (4\%)}
    \hfill \makecell{\textbf{Like} \\ 0 (0\%)} \hfill\null
\vspace{0.2em}
\\\Xhline{3\arrayrulewidth}
\midrule
 \textbf{Implicit ChatGPT Group} [\textbf{viewed}: 26 participants] \\\Xhline{3\arrayrulewidth}
\midrule
\scriptsize
\vspace{0.2em}
    \hfill \makecell{\textbf{Scroll Past} \\ 13 (50\%)} 
    \hfill \makecell{\textbf{Fact-check} \\ 7 (26.92\%)} 
    \hfill \makecell{\textbf{Block} \\ 2 (7.69\%)} 
    \hfill \makecell{\textbf{Report} \\ 2 (7.69\%)}
    \hfill \makecell{\textbf{Reply} \\ 1 (3.85\%)}
    \hfill \makecell{\textbf{Like} \\ 1 (3.85\%)} \hfill\null
\vspace{0.2em}
\\\Xhline{3\arrayrulewidth}
\bottomrule
\end{tabularx}
\end{table}

When the first videos were labeled as misinformation by TikTok and shown as such to the follow-up set of participants, there was a slight increase in perceiving the content as misinformation in both groups, as shown in Table \ref{tab:modvideo1-assessment}. While seven participants in the the explicit ChatGPT group (28\%) said they were ``\textit{not too knowledgeable on those herbs, but it seems accurate since it is ChatGPT}'' [\textbf{P105FM30}], four (16\%) said they ``\textit{don’t believe AI is the most reliable as it pulls from the information of the internet, which is full of misinformation; So, yes, in a way I believe this post could contain misinformation}'' [\textbf{P103FM20}]. Thirteen of the participants in the implicit ChatGPT group (52\%) said they ``\textit{cannot say with certainty that this post contains misinformation, but [they] would be very suspicious of it because of the lack of sources provided and the general language used}'' [\textbf{P138FL30}] and six (24\%) said ``\textit{this post contains misinformation because those herbs are not scientifically proven to induce safe abortion, and the video is exposing audiences to unsafe methods}'' [\textbf{P127FL20}]. 

\label{tab:modvideo1-assessment}
\begin{table}[htbp]
\renewcommand{\arraystretch}{1.3}
\footnotesize
\caption{Is Video \#1 Misinformation? [Labeled]}
\label{tab:modvideo1-assessment}
\centering
\aboverulesep=0ex 
   \belowrulesep=0ex 
\begin{tabularx}{\linewidth}{|Y|}
\Xhline{3\arrayrulewidth}
\toprule
 \textbf{Explicit ChatGPT Group} [\textbf{viewed}: 25 participants] \\\Xhline{3\arrayrulewidth}
\midrule
\footnotesize
\vspace{0.2em}
    \hfill \makecell{\textbf{Yes} \\ 4 (16\%)} 
    \hfill \makecell{\textbf{No} \\ 7 (28\%)} 
    \hfill \makecell{\textbf{Unsure} \\ 14 (56\%)} \hfill\null
\vspace{0.2em}
\\\Xhline{3\arrayrulewidth}
\midrule
 \textbf{Implicit ChatGPT Group} [\textbf{viewed}: 25 participants] \\\Xhline{3\arrayrulewidth}
\midrule
\footnotesize
\vspace{0.2em}
    \hfill \makecell{\textbf{Yes} \\ 6 (24\%)} 
    \hfill \makecell{\textbf{No} \\ 6 (24\%)} 
    \hfill \makecell{\textbf{Unsure} \\ 13 (52\%)} \hfill\null
\vspace{0.2em}
\\\Xhline{3\arrayrulewidth}
\bottomrule
\end{tabularx}
\end{table}

Though the majority of the participants were keen on scrolling pass the video ignoring of the misinformation label, as indicated in Table \ref{tab:modvideo1-action}, three participants in the explicit ChatGPT group (12\%)  said they ``\textit{would want to look up more information about abortifacient herbs and if there's any research or scientific studies on their effect on the human body}'' [\textbf{P102FL30}]. Two participants (8\%) said they ``\textit{would respond by commenting that ChatGPT isn't to be considered a medical or scientific source}'' [\textbf{P115ML30}] and one  participant said they would ``\textit{maybe block the account}'' [\textbf{P111MM30}]. Five participants in the implicit ChatGPT group (20\%) said they ``\textit{would go on Google to verify if there are herbs that cause abortion}'' [\textbf{P147FL30}]. The remaining participants were evenly split on their actions. Two said they ``\textit{would report the video and move on}'' [\textbf{P127FL20}], two said they ``\textit{would leave a comment to inform the person who posted it and others who might watch the video about how harmful the contents of the post are}'' [\textbf{P146FM20}], and two said they ``\textit{would probably like it}'' [\textbf{P141FL40}].

\begin{table}[htbp]
\renewcommand{\arraystretch}{1.3}
\footnotesize
\caption{What action would you take on Video \#1? [Labeled]}
\label{tab:modvideo1-action}
\centering
 \aboverulesep=0ex 
   \belowrulesep=0ex 
\begin{tabularx}{\linewidth}{|Y|}
\Xhline{3\arrayrulewidth}
\toprule
 \textbf{Explicit ChatGPT Group} [\textbf{viewed}: 25 participants] \\\Xhline{3\arrayrulewidth}
\midrule
\footnotesize
\vspace{0.2em}
    \hfill \makecell{\textbf{Scroll Past} \\ 18 (72\%)} 
    \hfill \makecell{\textbf{Fact-check} \\ 3 (12\%)} 
    \hfill \makecell{\textbf{Block} \\ 1 (4\%)} 
    \hfill \makecell{\textbf{Report} \\ 0 (0\%)}
    \hfill \makecell{\textbf{Reply} \\ 2 (8\%)}
    \hfill \makecell{\textbf{Like} \\ 1 (4\%)} \hfill\null
\vspace{0.2em}
\\\Xhline{3\arrayrulewidth}
\midrule
 \textbf{Implicit ChatGPT Group} [\textbf{viewed}: 25 participants] \\\Xhline{3\arrayrulewidth}
\midrule
\scriptsize
\vspace{0.2em}
    \hfill \makecell{\textbf{Scroll Past} \\ 14 (56\%)} 
    \hfill \makecell{\textbf{Fact-check} \\ 5 (20\%)} 
    \hfill \makecell{\textbf{Block} \\ 0 (0\%)} 
    \hfill \makecell{\textbf{Report} \\ 2 (8\%)}
    \hfill \makecell{\textbf{Reply} \\ 2 (8\%)}
    \hfill \makecell{\textbf{Like} \\ 2 (8\%)} \hfill\null
\vspace{0.2em}
\\\Xhline{3\arrayrulewidth}
\bottomrule
\end{tabularx}
\end{table}

\subsection{Tell me about herbs for abortion}
The second set of TikTok videos, shown in Appendix \ref{sec:video2}, were created using the prompt ``\textit{Tell me about herbs for abortion}''. As indicated in Table \ref{tab:video2-assessment}, most participants in both groups felt that these videos were not misinformation. Seventeen participants in the explicit ChatGPT group (65.38\%) stated the information in the video seems to be factual because ``\textit{it at least encourages consulting trained medical professionals rather than the internet, which I think is the correct response in that type of situation}'' [\textbf{P23FM40}]. Eleven of the participants in the implicit ChatGPT group (45.83\%) felt the video was not misinformation, adding that ``\textit{it's just somebody posting some FYI}'' [\textbf{P97FM40}]. 

Eight of participants in the explicit ChatGPT group (30.77\%) stated they ``\textit{can't elaborate on the validity of the chat bot due to it just being good at writing and not realizing what it is actually writing}'' [\textbf{P44ML20}]. Nine of the implicit ChatGPT group participants (37.5\%) also were unsure, but felt the information ``\textit{would be better coming from someone that was an expert}'' [\textbf{P66FM20}]. Only one participant from the explicit ChatGPT group thought the video ``\textit{looks as though it contains misinformation regarding abortions in this form}'' [\textbf{P34MR30}]. Four of the participants in the implicit ChatGPT group (16.67\%) thought the video contained misinformation because ``\textit{this TikTok is an advertisement for a product and is therefore biased toward selling vs informing}'' [\textbf{P92FL40}].

\label{tab:video2-assessment}
\begin{table}[htbp]
\renewcommand{\arraystretch}{1.3}
\footnotesize
\caption{Is Video \#2 Misinformation?}
\label{tab:video2-assessment}
\centering
\aboverulesep=0ex 
   \belowrulesep=0ex 
\begin{tabularx}{\linewidth}{|Y|}
\Xhline{3\arrayrulewidth}
\toprule
 \textbf{Explicit ChatGPT Group} [\textbf{viewed}: 26 participants] \\\Xhline{3\arrayrulewidth}
\midrule
\footnotesize
\vspace{0.2em}
    \hfill \makecell{\textbf{Yes} \\ 1 (3.85\%)} 
    \hfill \makecell{\textbf{No} \\ 17 (65.38\%)} 
    \hfill \makecell{\textbf{Unsure} \\ 8 (30.77\%)} \hfill\null
\vspace{0.2em}
\\\Xhline{3\arrayrulewidth}
\midrule
 \textbf{Implicit ChatGPT Group} [\textbf{viewed}: 24 participants] \\\Xhline{3\arrayrulewidth}
\midrule
\footnotesize
\vspace{0.2em}
    \hfill \makecell{\textbf{Yes} \\ 4 (16.67\%)} 
    \hfill \makecell{\textbf{No} \\ 11 (45.83\%)} 
    \hfill \makecell{\textbf{Unsure} \\ 9 (37.5\%)} \hfill\null
\vspace{0.2em}
\\\Xhline{3\arrayrulewidth}
\bottomrule
\end{tabularx}
\end{table}

Again, the majority of the participants in both groups, as shown in Table \ref{tab:video2-action}, stated they ``\textit{would honestly probably scroll past it, although the functionality of ChatGPT is impressive}'' [\textbf{P4MM30}]. Four of the participants in the explicit ChatGPT group (15.38\%) said they would ``\textit{like the post because answers the question, however it also explains how dangerous it would be to use.}'' [\textbf{P41FL30}]. Three of participants in the explicit ChatGPT group (11.54\%) said they ``\textit{would probably read the comments to see what others said}'' [\textbf{P36FL30}], one said they ``\textit{would share it}'' [\textbf{P22FM40}], and one said they ``\textit{would respond with skepticism and confusion}'' [\textbf{P35FL30}]. Five of participants in the implicit ChatGPT group (20.83\%) said they ``\textit{would research for themselves what herbs can be used for abortion}'' [\textbf{P61FL30}], one said ``\textit{this is a good video that I would feel comfortable sharing}'' [\textbf{P67FL40}] and one said ``\textit{if this post came on my feed I would flag it}'' [\textbf{P81ML40}].

\begin{table}[htbp]
\renewcommand{\arraystretch}{1.3}
\footnotesize
\caption{What action would you take on Video \#2?}
\label{tab:video2-action}
\centering
 \aboverulesep=0ex 
    \belowrulesep=0ex 
\begin{tabularx}{\linewidth}{|Y|}
\Xhline{3\arrayrulewidth}
\toprule
 \textbf{Explicit ChatGPT Group} [\textbf{viewed}: 26 participants] \\\Xhline{3\arrayrulewidth}
\midrule
\scriptsize
\vspace{0.2em}
    \hfill \makecell{\textbf{Scroll Past} \\ 17 (65.38\%)} 
    \hfill \makecell{\textbf{Fact-check} \\ 3 (11.54\%)} 
    \hfill \makecell{\textbf{Share} \\ 1 (3.85\%)} 
    \hfill \makecell{\textbf{Report} \\ 0 (0\%)}
    \hfill \makecell{\textbf{Reply} \\ 1 (3.85\%)}
    \hfill \makecell{\textbf{Like} \\ 4 (15.38\%)} \hfill\null
\vspace{0.2em}
\\\Xhline{3\arrayrulewidth}
\midrule
 \textbf{Implicit ChatGPT Group} [\textbf{viewed}: 24 participants] \\\Xhline{3\arrayrulewidth}
\midrule
\scriptsize
\vspace{0.2em}
    \hfill \makecell{\textbf{Scroll Past} \\ 16 (66.66\%)} 
    \hfill \makecell{\textbf{Fact-check} \\ 5 (20.83\%)} 
    \hfill \makecell{\textbf{Share} \\ 1 (4.17\%)} 
    \hfill \makecell{\textbf{Report} \\ 1 (4.17\%)}
    \hfill \makecell{\textbf{Reply} \\ 1 (4.17\%)}
    \hfill \makecell{\textbf{Like} \\ 0 (0\%)} \hfill\null
\vspace{0.2em}
\\\Xhline{3\arrayrulewidth}
\bottomrule
\end{tabularx}
\end{table}
The misinformation label did not cause any noticeable effect in the perception of the videos for both groups, as indicated in Table \ref{tab:modvideo2-assessment}. Nine participants in the explicit ChatGPT group (36\%) said the video ``\textit{seems like helpful information without any misinformation}'' [\textbf{P105FM30}] but two (8\%) said they felt it was misinformation because ``\textit{some of the herbs don't cause abortion}'' [\textbf{P124FM40}]. Eleven of the participants in the implicit ChatGPT group said they ``\textit{don’t necessarily think they are spreading misinformation, but [they] don't feel they are giving enough truth in their post either}'' [\textbf{P128FL50}]. Ten participants in this group (40\%) said ``\textit{the post does not include any misinformation and is accurate}'' [\textbf{P147FL30}], and four (16\%) said they ``\textit{believe this post does contain misinformation because herbs are not a very common form of abortion healthcare}'' [\textbf{P138FL30}].

\label{tab:modvideo2-assessment}
\begin{table}[htbp]
\renewcommand{\arraystretch}{1.3}
\footnotesize
\caption{Is Video \#2 Misinformation? [Labeled]}
\label{tab:modvideo2-assessment}
\centering
\aboverulesep=0ex 
   \belowrulesep=0ex 
\begin{tabularx}{\linewidth}{|Y|}
\Xhline{3\arrayrulewidth}
\toprule
 \textbf{Explicit ChatGPT Group} [\textbf{viewed}: 25 participants] \\\Xhline{3\arrayrulewidth}
\midrule
\footnotesize
\vspace{0.2em}
    \hfill \makecell{\textbf{Yes} \\ 2 (8\%)} 
    \hfill \makecell{\textbf{No} \\ 9 (36\%)} 
    \hfill \makecell{\textbf{Unsure} \\ 14 (56\%)} \hfill\null
\vspace{0.2em}
\\\Xhline{3\arrayrulewidth}
\midrule
 \textbf{Implicit ChatGPT Group} [\textbf{viewed}: 25 participants] \\\Xhline{3\arrayrulewidth}
\midrule
\footnotesize
\vspace{0.2em}
    \hfill \makecell{\textbf{Yes} \\ 4 (16\%)} 
    \hfill \makecell{\textbf{No} \\ 10 (40\%)} 
    \hfill \makecell{\textbf{Unsure} \\ 11 (44\%)} \hfill\null
\vspace{0.2em}
\\\Xhline{3\arrayrulewidth}
\bottomrule
\end{tabularx}
\end{table}

Watching the video and scrolling past was still the main engagement strategy for both groups, as shown in Table \ref{tab:modvideo2-action}, though three participants in the explicit ChatGPT group (12\%) said they ``\textit{would take in the information and read the comments and likely look up more about those types of herbs and their practice in the past for use in at home abortions}'' [\textbf{P102FL30}]. Another three (12\%) said they ``\textit{would respond by commenting that ChatGPT first paragraph should be taken seriously, since the use of such herbs could be fatal or very damaging}'' [\textbf{P115ML30}]. One participant actually pointed to the misinformation label attached by TikTok, saying they ``\textit{like that the TikTok stresses how dangerous they are though, so I might like the post if I saw it}'' [\textbf{P123FL30}]. In the implicit ChatGPT group, four of the participants (16\%) said they ``\textit{would like the post since it does not include any misinformation and is accurate}'' [\textbf{P147FL30}]. 

\begin{table}[htbp]
\renewcommand{\arraystretch}{1.3}
\footnotesize
\caption{What action would you take on Video \#2? [Labeled]}
\label{tab:modvideo2-action}
\centering
 \aboverulesep=0ex 
   \belowrulesep=0ex 
\begin{tabularx}{\linewidth}{|Y|}
\Xhline{3\arrayrulewidth}
\toprule
 \textbf{Explicit ChatGPT Group} [\textbf{viewed}: 25 participants] \\\Xhline{3\arrayrulewidth}
\midrule
\footnotesize
\vspace{0.2em}
    \hfill \makecell{\textbf{Scroll Past} \\ 18 (72\%)} 
    \hfill \makecell{\textbf{Fact-check} \\ 3 (12\%)} 
    \hfill \makecell{\textbf{Block} \\ 0 (0\%)} 
    \hfill \makecell{\textbf{Report} \\ 0 (0\%)}
    \hfill \makecell{\textbf{Reply} \\ 3 (12\%)}
    \hfill \makecell{\textbf{Like} \\ 1 (4\%)} \hfill\null
\vspace{0.2em}
\\\Xhline{3\arrayrulewidth}
\midrule
 \textbf{Implicit ChatGPT Group} [\textbf{viewed}: 25 participants] \\\Xhline{3\arrayrulewidth}
\midrule
\scriptsize
\vspace{0.2em}
    \hfill \makecell{\textbf{Scroll Past} \\ 18 (72\%)} 
    \hfill \makecell{\textbf{Fact-check} \\ 1 (4\%)} 
    \hfill \makecell{\textbf{Block} \\ 1 (4\%)} 
    \hfill \makecell{\textbf{Report} \\ 1 (4\%)}
    \hfill \makecell{\textbf{Reply} \\ 0 (0\%)}
    \hfill \makecell{\textbf{Like} \\ 4 (16\%)} \hfill\null
\vspace{0.2em}
\\\Xhline{3\arrayrulewidth}
\bottomrule
\end{tabularx}
\end{table}

\subsection{Tell me all of the facts about pennyroyal}
The third set of videos, shown in Appendix \ref{sec:video3}, were created using the prompt ``\textit{Tell me all of the facts about pennyroyal}.'' As shown in Table \ref{tab:video3-assessment}, 16 (66.67\%) of participants in the explicit ChatGPT group stated they ``\textit{do not know enough about Pennyroyal to know if it is misinformation but I do know not to trust what ChatGPT says}'' [\textbf{P9FL30}]. Six  participants in this group (25\%) stated ``\textit{it is unlikely the information is misinformation as it appears to have pulled up a definition from the internet}'' [\textbf{P19ML30}] and two (8.33\%) ``\textit{stated that feel that leaving information out of the description is a form of misinformation and it should have been included for someone to have a full picture of the facts}'' [\textbf{P13FL30}]. The implicit ChatGPT group participants were mostly split between saying they ``\textit{have no idea if this is misinformation}'' [\textbf{P58FMNR}] and believing ``\textit{the information is probably correct list the potential risks associated with the herb}'' [\textbf{P52NCL30}]. The remaining four participants in this group (16\%)  thought ``\textit{this post contains misinformation, as people sometimes just post information they know from their grandmas for example, or from someone else who doesn't have any studies in life, and that can make people feel confused}'' [\textbf{P63FM30}].

\label{tab:video3-assessment}
\begin{table}[htbp]
\renewcommand{\arraystretch}{1.3}
\footnotesize
\caption{Is Video \#3 Misinformation?}
\label{tab:video3-assessment}
\centering
\aboverulesep=0ex 
   \belowrulesep=0ex 
\begin{tabularx}{\linewidth}{|Y|}
\Xhline{3\arrayrulewidth}
\toprule
 \textbf{Explicit ChatGPT Group} [\textbf{viewed}: 24 participants] \\\Xhline{3\arrayrulewidth}
\midrule
\footnotesize
\vspace{0.2em}
    \hfill \makecell{\textbf{Yes} \\ 2 (8.33\%)} 
    \hfill \makecell{\textbf{No} \\ 6 (25\%)} 
    \hfill \makecell{\textbf{Unsure} \\ 16 (66.67\%)} \hfill\null
\vspace{0.2em}
\\\Xhline{3\arrayrulewidth}
\midrule
 \textbf{Implicit ChatGPT Group} [\textbf{viewed}: 25 participants] \\\Xhline{3\arrayrulewidth}
\midrule
\footnotesize
\vspace{0.2em}
    \hfill \makecell{\textbf{Yes} \\ 4 (16\%)} 
    \hfill \makecell{\textbf{No} \\ 10 (40\%)} 
    \hfill \makecell{\textbf{Unsure} \\ 11 (44\%)} \hfill\null
\vspace{0.2em}
\\\Xhline{3\arrayrulewidth}
\bottomrule
\end{tabularx}
\end{table}

About half of the participants in each group shown in Table \ref{tab:video3-action} said they would watch and scroll past the videos. Seven of the participants in the explicit ChatGPT group (29.17\%) said they ``\textit{would do more research about this herb because it sounds interesting and helpful}'' [\textbf{P24FL30}]. Two participants that said they would reply to `\textit{comment that pennyroyal should not be ingested.. ever}'' [\textbf{P18FL50}]. The remaining participant said they ``\textit{may attempt to report it as misinformation}'' [\textbf{P13FL30}]. In the implicit ChatGPT group, eight participants (32\%) said they ``\textit{would go to a legitimate resource on herbs and their medicinal effects if I had an interest in the information here}'' [\textbf{P91NCL60}] and two said they ``\textit{would block the account to prevent any more posts from appearing on my for you page again}'' [\textbf{P82FL30}]. One participant was keen on ``\textit{reporting it}'' [\textbf{P74FM30}], and one said they would ``\textit{maybe like it so content like it could come on my feed}'' [\textbf{P52NCL30}].

\begin{table}[htbp]
\renewcommand{\arraystretch}{1.3}
\footnotesize
\caption{What action would you take on Video \#3?}
\label{tab:video3-action}
\centering
 \aboverulesep=0ex 
    \belowrulesep=0ex 
\begin{tabularx}{\linewidth}{|Y|}
\Xhline{3\arrayrulewidth}
\toprule
 \textbf{Explicit ChatGPT Group} [\textbf{viewed}: 24 participants] \\\Xhline{3\arrayrulewidth}
\midrule
\footnotesize
\vspace{0.2em}
    \hfill \makecell{\textbf{Scroll Past} \\ 14 (58.33\%)} 
    \hfill \makecell{\textbf{Fact-check} \\ 7 (29.17\%)} 
    \hfill \makecell{\textbf{Block} \\ 0 (0\%)} 
    \hfill \makecell{\textbf{Report} \\ 1 (4.17\%)}
    \hfill \makecell{\textbf{Reply} \\ 2 (8.33\%)}
    \hfill \makecell{\textbf{Like} \\ 0 (0\%)} \hfill\null
\vspace{0.2em}
\\\Xhline{3\arrayrulewidth}
\midrule
 \textbf{Implicit ChatGPT Group} [\textbf{viewed}: 25 participants] \\\Xhline{3\arrayrulewidth}
\midrule
\footnotesize
\vspace{0.2em}
    \hfill \makecell{\textbf{Scroll Past} \\ 13 (52\%)} 
    \hfill \makecell{\textbf{Fact-check} \\ 8 (32\%)} 
    \hfill \makecell{\textbf{Block} \\ 2 (8\%)} 
    \hfill \makecell{\textbf{Report} \\ 1 (4\%)}
    \hfill \makecell{\textbf{Reply} \\ 0 (0\%)}
    \hfill \makecell{\textbf{Like} \\ 1 (4\%)} \hfill\null
\vspace{0.2em}
\\\Xhline{3\arrayrulewidth}
\bottomrule
\end{tabularx}
\end{table}
The misinformation label for the third set of videos also did not have any effect on the perceived accuracy of both the explicit and implicit ChatGPT responses, as shown in Table \ref{tab:modvideo3-assessment}. In fact, more than half of participants (56\%) in the explicit ChatGPT group thought that the video not misinformation because the ``\textit{information was very detailed and specific which leads me to believe it’s likely accurate}'' [\textbf{P101FL20}]. Participant \textbf{P108MR30} added that ``\textit{it seems like this info came straight from Wikipedia though so my guess is it's likely accurate}.'' Nine of participants in this group (36\%) were uncertain as they ``\textit{know little about pennyroyals}'' [\textbf{P122MM30}] and two (8\%) said ``\textit{the post contains misinformation with regards to the fact that pennyroyal is toxic regardless of dosage and adverse effects such as vomiting and dizziness can occur after ingestion of doses less than 10 milliliters}'' [\textbf{P119NCL30}]. Fifteen of the participants from the implicit ChatGPT group (60\%) said they are ``\textit{not sure if it contains misinformation but it sounds more knowledgeable than the others so I would look into it; The more detail there are, the more convincing it is}'' [\textbf{P138FL30}]. Nine participants in this group (36\%) said ``\textit{the idea of misinformation being in this is extremely low all thanks to the fact that the creator made it seem like they did their research}'' [\textbf{P136MR30}]. 

\label{tab:modvideo3-assessment}
\begin{table}[htbp]
\renewcommand{\arraystretch}{1.3}
\footnotesize
\caption{Is Video \#3 Misinformation? [Labeled]}
\label{tab:modvideo3-assessment}
\centering
\aboverulesep=0ex 
   \belowrulesep=0ex 
\begin{tabularx}{\linewidth}{|Y|}
\Xhline{3\arrayrulewidth}
\toprule
 \textbf{Explicit ChatGPT Group} [\textbf{viewed}: 25 participants] \\\Xhline{3\arrayrulewidth}
\midrule
\footnotesize
\vspace{0.2em}
    \hfill \makecell{\textbf{Yes} \\ 2 (8\%)} 
    \hfill \makecell{\textbf{No} \\ 14 (56\%)} 
    \hfill \makecell{\textbf{Unsure} \\ 9 (36\%)} \hfill\null
\vspace{0.2em}
\\\Xhline{3\arrayrulewidth}
\midrule
 \textbf{Implicit ChatGPT Group} [\textbf{viewed}: 25 participants] \\\Xhline{3\arrayrulewidth}
\midrule
\footnotesize
\vspace{0.2em}
    \hfill \makecell{\textbf{Yes} \\ 1 (4\%)} 
    \hfill \makecell{\textbf{No} \\ 9 (36\%)} 
    \hfill \makecell{\textbf{Unsure} \\ 15 (60\%)} \hfill\null
\vspace{0.2em}
\\\Xhline{3\arrayrulewidth}
\bottomrule
\end{tabularx}
\end{table}

As shown in Table \ref{tab:modvideo3-action}, beyond the usual ``scroll past,'' five participants in the explicit ChatGPT group (20\%) said they ``\textit{would probably look up the herb and read about it}'' [\textbf{P113FM40}]. Two participants (8\%) said they ``\textit{might like the post, but I wouldn't share it because I would be worried about someone trying to do something unsafe with the information If they were desperate}'' [\textbf{P123FL30}], and one said they ``\textit{would would respond by not believing the information}'' [\textbf{P121FL40}]. In the implicit ChatGPT group, similarly, nine of the participants (36\%) said they ``\textit{would read it and maybe look up that herb to see if that is true}'' [\textbf{P143FL30}], and three (12\%) said they ``\textit{would like the post}'' [\textbf{P146FM20}].

\begin{table}[htbp]
\renewcommand{\arraystretch}{1.3}
\footnotesize
\caption{What action would you take on Video \#3? [Labeled]}
\label{tab:modvideo3-action}
\centering
 \aboverulesep=0ex 
   \belowrulesep=0ex 
\begin{tabularx}{\linewidth}{|Y|}
\Xhline{3\arrayrulewidth}
\toprule
 \textbf{Explicit ChatGPT Group} [\textbf{viewed}: 25 participants] \\\Xhline{3\arrayrulewidth}
\midrule
\footnotesize
\vspace{0.2em}
    \hfill \makecell{\textbf{Scroll Past} \\ 17 (68\%)} 
    \hfill \makecell{\textbf{Fact-check} \\ 5 (20\%)} 
    \hfill \makecell{\textbf{Block} \\ 0 (0\%)} 
    \hfill \makecell{\textbf{Report} \\ 0 (0\%)}
    \hfill \makecell{\textbf{Reply} \\ 1 (4\%)}
    \hfill \makecell{\textbf{Like} \\ 2 (8\%)} \hfill\null
\vspace{0.2em}
\\\Xhline{3\arrayrulewidth}
\midrule
 \textbf{Implicit ChatGPT Group} [\textbf{viewed}: 25 participants] \\\Xhline{3\arrayrulewidth}
\midrule
\scriptsize
\vspace{0.2em}
    \hfill \makecell{\textbf{Scroll Past} \\ 13 (52\%)} 
    \hfill \makecell{\textbf{Fact-check} \\ 9 (36\%)} 
    \hfill \makecell{\textbf{Block} \\ 0 (0\%)} 
    \hfill \makecell{\textbf{Report} \\ 0 (0\%)}
    \hfill \makecell{\textbf{Reply} \\ 0 (0\%)}
    \hfill \makecell{\textbf{Like} \\ 3 (12\%)} \hfill\null
\vspace{0.2em}
\\\Xhline{3\arrayrulewidth}
\bottomrule
\end{tabularx}
\end{table}

\subsection{Write a rap about pennyroyal}
The fourth set of videos, shown in Appendix \ref{sec:video4}, were created using the prompt ``\textit{Write a rap about pennyroyal}''. As shown in Table \ref{tab:video4-assessment}, most participants were unsure if this video was misinformation. Sixteen of the participants in the explicit ChatGPT group (61.54\%) stated they ``\textit{can't tell if the post has any misinformation as it is a rap and most songs aren't made for educational purposes}'' [\textbf{P35FL30}]. Participant \textbf{P23FM40} added that ``\textit{some of the information in it might be untrue or misleading, but since it is presented in the sort of goofy rap form, I'm not sure this is meant to be taken seriously}.'' 

One of the participants, \textbf{P9FL30}, pointed that ``\textit{ChatGPT is not a trusted source}''. Nine participants in the explicit ChatGPT group (34.62\%) said ``\textit{it doesn't appear there is any misinformation; It looked like it was just a made up rap song}'' [\textbf{P21FR50}], and one said ``\textit{don't believe it technically contains misinformation as poetry isn't exactly making claims and could be fictional}'' [\textbf{P38MR30}]. In the implicit ChatGPT group, 12 participants (52.17\%) said they ``\textit{don't know if this TikTok has any misinformation}'' [\textbf{P63FM30}] and three (13.04\%) said ``\textit{the information didn’t express the dangers of the herb strongly enough, so it is a misinformation through omission}'' [\textbf{P65FL61+}]. Eight participants (34.78\%) said ``\textit{it has very bad grammar and etiquette which leads me to believe it contains misinformation}'' [\textbf{P70FM40}] and commented that the ``\textit{lyrics sound childish, does not seem accurate.}'' [\textbf{P84FL30}].

\begin{table}[htbp]
\renewcommand{\arraystretch}{1.3}
\footnotesize
\caption{Is Video \#4 Misinformation?}
\label{tab:video4-assessment}
\centering
\aboverulesep=0ex 
   \belowrulesep=0ex 
\begin{tabularx}{\linewidth}{|Y|}
\Xhline{3\arrayrulewidth}
\toprule
 \textbf{Explicit ChatGPT Group} [\textbf{viewed}: 26 participants] \\\Xhline{3\arrayrulewidth}
\midrule
\footnotesize
\vspace{0.2em}
    \hfill \makecell{\textbf{Yes} \\ 1 (3.84\%)} 
    \hfill \makecell{\textbf{No} \\ 9 (34.62\%)} 
    \hfill \makecell{\textbf{Unsure} \\ 16 (61.54\%)} \hfill\null
\vspace{0.2em}
\\\Xhline{3\arrayrulewidth}
\midrule
 \textbf{Implicit ChatGPT Group} [\textbf{viewed}: 23 participants] \\\Xhline{3\arrayrulewidth}
\midrule
\footnotesize
\vspace{0.2em}
    \hfill \makecell{\textbf{Yes} \\ 8 (34.78\%)} 
    \hfill \makecell{\textbf{No} \\ 3 (13.04\%)} 
    \hfill \makecell{\textbf{Unsure} \\ 12 (52.17\%)} \hfill\null
\vspace{0.2em}
\\\Xhline{3\arrayrulewidth}
\bottomrule
\end{tabularx}
\end{table}

As indicated in Table \ref{tab:video4-action}, 17 participants in the explicit ChatGPT group (65.38\%) said they ``\textit{would scroll past this}'' [\textbf{P31FL40}] and six (23.08\%) ``\textit{would be intrigued to validate the claims and research it externally}'' [\textbf{P38MR30}]. One of the participants said they ``\textit{would report the video for encouraging the use of an herb without providing any actual support for why it would work}'' [\textbf{P25FM30}]. One participant said the would ``\textit{probably comment on it and send it to a friend because it's interesting that AI generates that}'' [\textbf{P34MR30}], and one said they ``\textit{would like the video}'' [\textbf{P21FR50}]. 

In the implicit ChatGPT group, 16 participants (69.57\%) said they they ``\textit{would not trust something in a rhyme, as it seems like a spell casting, its creepy}'' [\textbf{P78FL20}]. Four participants (17.39\%) said they ``\textit{would comment on how they know what they are implying by the caption even if the poem doesn't explicitly state it and it can harm someone; I'd make a Facebook post telling people not to listen to this; It can harm}'' [\textbf{P100NC40}]. Two participants said they ``\textit{probably would be interested in finding out more about the herb and start [their] own research}'' [\textbf{P57FA30}] and one said they ``\textit{would probably report this video because it is most certainly a scam}'' [\textbf{P76MM30}].

\begin{table}[htbp]
\renewcommand{\arraystretch}{1.3}
\footnotesize
\caption{What action would you take on Video \#4?}
\label{tab:video4-action}
\centering
\aboverulesep=0ex 
   \belowrulesep=0ex 
\begin{tabularx}{\linewidth}{|Y|}
\Xhline{3\arrayrulewidth}
\toprule
 \textbf{Explicit ChatGPT Group} [\textbf{viewed}: 26 participants] \\\Xhline{3\arrayrulewidth}
\midrule
\footnotesize
\vspace{0.2em}
    \hfill \makecell{\textbf{Scroll Past} \\ 17 (65.38\%)} 
    \hfill \makecell{\textbf{Fact-check} \\ 6 (23.08\%)} 
    \hfill \makecell{\textbf{Block} \\ 0 (0\%)} 
    \hfill \makecell{\textbf{Report} \\ 1 (3.85\%)}
    \hfill \makecell{\textbf{Reply} \\ 1 (3.85\%)}
    \hfill \makecell{\textbf{Like} \\ 1 (3.85\%)} \hfill\null
\vspace{0.2em}
\\\Xhline{3\arrayrulewidth}
\midrule
 \textbf{Implicit ChatGPT Group} [\textbf{viewed}: 23 participants] \\\Xhline{3\arrayrulewidth}
\midrule
\footnotesize
\vspace{0.2em}
    \hfill \makecell{\textbf{Scroll Past} \\ 16 (69.57\%)} 
    \hfill \makecell{\textbf{Fact-check} \\ 2 (8.7\%)} 
    \hfill \makecell{\textbf{Block} \\ 0 (0\%)} 
    \hfill \makecell{\textbf{Report} \\ 1 (4.35\%)}
    \hfill \makecell{\textbf{Reply} \\ 4 (17.39\%)}
    \hfill \makecell{\textbf{Like} \\ 0 (0\%)} \hfill\null
\vspace{0.2em}
\\\Xhline{3\arrayrulewidth}
\bottomrule
\end{tabularx}
\end{table}
The label paired with the mobile-only access of the videos made more participants in the explicit ChatGPT group beleive the rap lyrics were misinformation, as shown in \ref{tab:modvideo4-assessment}. Their impression was that ``\textit{the rap could be dangerous since it encourages its use even if it’s toxic, without recommending talking to a health care professional}'' [\textbf{P105FM30}], feeling that ``\textit{it would make [them] smile a bit, but it can be seen as misinformation purely because the rap makes it seem like some crazy cool herb that'll make you feel great and it doesn't warn against taking too much}'' [\textbf{P118FL30}]. In the implicit ChatGPT group, 11 participants (44\%) said they ``\textit{would be suspicious of this post; I don't like that it's written like an ad because it seems like it's trying to sell me something rather than to inform me; It could contain misinformation for that reason}'' [\textbf{P140FL20}]. Eight of the participants in this group (32\%) said ``\textit{the post doesn't make much sense, so I wouldn't trust it; I'd assume that there would be misinformation in the post due to it not making sense}'' [\textbf{P148FL40}]. Six participants (24\%) said ``\textit{the rhyme didn't seem to contain any real info so no misinformation to my eye}'' [\textbf{P126ML40}]. 

\label{tab:modvideo4-assessment}
\begin{table}[htbp]
\renewcommand{\arraystretch}{1.3}
\footnotesize
\caption{Is Video \#4 Misinformation? [Labeled]}
\label{tab:modpost7-assessment}
\centering
\aboverulesep=0ex 
   \belowrulesep=0ex 
\begin{tabularx}{\linewidth}{|Y|}
\Xhline{3\arrayrulewidth}
\toprule
 \textbf{Explicit ChatGPT Group} [\textbf{viewed}: 25 participants] \\\Xhline{3\arrayrulewidth}
\midrule
\footnotesize
\vspace{0.2em}
    \hfill \makecell{\textbf{Yes} \\ 7 (28\%)} 
    \hfill \makecell{\textbf{No} \\ 7 (28\%)} 
    \hfill \makecell{\textbf{Unsure} \\ 11 (44\%)} \hfill\null
\vspace{0.2em}
\\\Xhline{3\arrayrulewidth}
\midrule
 \textbf{Implicit ChatGPT Group} [\textbf{viewed}: 25 participants] \\\Xhline{3\arrayrulewidth}
\midrule
\footnotesize
\vspace{0.2em}
    \hfill \makecell{\textbf{Yes} \\ 8 (32\%)} 
    \hfill \makecell{\textbf{No} \\ 6 (24\%)} 
    \hfill \makecell{\textbf{Unsure} \\ 11 (44\%)} \hfill\null
\vspace{0.2em}
\\\Xhline{3\arrayrulewidth}
\bottomrule
\end{tabularx}
\end{table}

As shown in Table \ref{tab:modvideo4-action}, most of the explicit ChatGPT group participants had mixed feeling about ChatGPT creating lyrical content, commenting that ``\textit{it's kind of funny to see AI make a rap about abortion}'' [\textbf{P108MR30}]. The remaining two participants were either going to``\textit{ research then block if false}'' [\textbf{P106MM40}] or ``\textit{would respond by not believing the information}'' [\textbf{P121FL40}]. Nineteen of the of the implicit ChatGPT group (76\%)  said they think ``\textit{think this post is odd}'' [\textbf{P121FL40}]. The remaining two participants in this group said  they ``\textit{would do my own research via Google}'' [\textbf{P130FA60}] and comment on the video to ``\textit{encourage others to do their research before trying pennyroyal}'' [\textbf{P146FM20}].

\begin{table}[htbp]
\renewcommand{\arraystretch}{1.3}
\footnotesize
\caption{What action would you take on Video \#4? [Labeled]}
\label{tab:modvideo4-action}
\centering
 \aboverulesep=0ex 
   \belowrulesep=0ex 
\begin{tabularx}{\linewidth}{|Y|}
\Xhline{3\arrayrulewidth}
\toprule
 \textbf{Explicit ChatGPT Group} [\textbf{viewed}: 25 participants] \\\Xhline{3\arrayrulewidth}
\midrule
\footnotesize
\vspace{0.2em}
    \hfill \makecell{\textbf{Scroll Past} \\ 23 (92\%)} 
    \hfill \makecell{\textbf{Fact-check} \\ 1 (4\%)} 
    \hfill \makecell{\textbf{Block} \\ 0 (0\%)} 
    \hfill \makecell{\textbf{Report} \\ 0 (0\%)}
    \hfill \makecell{\textbf{Reply} \\ 1 (4\%)}
    \hfill \makecell{\textbf{Like} \\ 0 (0\%)} \hfill\null
\vspace{0.2em}
\\\Xhline{3\arrayrulewidth}
\midrule
 \textbf{Implicit ChatGPT Group} [\textbf{viewed}: 25 participants] \\\Xhline{3\arrayrulewidth}
\midrule
\scriptsize
\vspace{0.2em}
    \hfill \makecell{\textbf{Scroll Past} \\ 19 (76\%)} 
    \hfill \makecell{\textbf{Fact-check} \\ 2 (8\%)} 
    \hfill \makecell{\textbf{Block} \\ 0 (0\%)} 
    \hfill \makecell{\textbf{Report} \\ 0 (0\%)}
    \hfill \makecell{\textbf{Reply} \\ 2 (8\%)}
    \hfill \makecell{\textbf{Like} \\ 2 (8\%)} \hfill\null
\vspace{0.2em}
\\\Xhline{3\arrayrulewidth}
\bottomrule
\end{tabularx}
\end{table}

\section{Results: Reception} \label{sec:results-reception}

After viewing the videos, participants in both groups in each study were asked about their prior experience with chatbots or language models and their reception of them. Most participants indicated they had very limited experience with chatbots other than using them for online customer service, but had heard about them in the news. We categorized participants opinions of chatbots and language models based on their tone of their response, namely positive, negative, hesitant, or absent of an opinion.

Participants in the explicit ChatGPT group generally had a more positive view of chatbots, as indicated in Table \ref{tab:reception}, and lauded the  ``\textit{generally knowledgeability} [\textbf{\textbf{P118FL30}}]. Fifteen  participants overall had a negative opinion of chatbots included noting they ``\textit{don't trust bots to give them information and would want to research from credible sources before trusting a bot}'' [\textbf{P40FL30}]. These participants thought that ``\textit{chatbots or language models can be dangerous and are not reliable resources for information because they seemingly provide people with a sense of entertainment}'' [\textbf{P13FL30}]. The participants with a hesitant response pointed to the credibility of chatbots' responses, stating ``\textit{they can be useful for a quick glimpse into information, but should not be taken as an absolute; They facilitate a start in research, but are not and should not be taken as scientific proof}'' [\textbf{P19ML30}].

\begin{table}[htbp]
\renewcommand{\arraystretch}{1.3}
\footnotesize
\caption{Opinion of Chatbots or Language Models}
\label{tab:reception}
\centering
\aboverulesep=0ex 
   \belowrulesep=0ex 
\begin{tabularx}{\linewidth}{|Y|}
\Xhline{3\arrayrulewidth}
\toprule
 \textbf{Explicit ChatGPT Group} [\textbf{viewed}: 75 participants] \\\Xhline{3\arrayrulewidth}
\midrule
\footnotesize
\vspace{0.2em}
    \hfill \makecell{\textbf{Positive} \\ 25 (33.33\%)} 
    \hfill \makecell{\textbf{Negative} \\ 15 (20\%)} 
    \hfill \makecell{\textbf{Hesitant} \\ 30 (40\%)}
    \hfill \makecell{\textbf{Neutral} \\ 5 (6.67\%)}\hfill\null
\vspace{0.2em}
\\\Xhline{3\arrayrulewidth}
\midrule
 \textbf{Implicit ChatGPT Group} [\textbf{viewed}: 75 participants] \\\Xhline{3\arrayrulewidth}
\midrule
\footnotesize
\vspace{0.2em}
    \hfill \makecell{\textbf{Positive} \\ 19 (25.33\%)} 
    \hfill \makecell{\textbf{Negative} \\ 11 (14.67\%)} 
    \hfill \makecell{\textbf{Hesitant} \\ 37 (49.33\%)}
    \hfill \makecell{\textbf{Neutral} \\ 8 (10.67\%)}\hfill\null
\vspace{0.2em}
\\\Xhline{3\arrayrulewidth}
\bottomrule
\end{tabularx}
\end{table}

Participant \textbf{P119NCL30} commented that ``\textit{language models are very intriguing and can be useful as a resource, but I am wary of the information that it returns when given certain prompts; Chatbots can output fabricated information such as citing nonexistent studies, which can be very dangerous if taken at face value and spread to other people}''. Another participant, \textbf{P125MR30} felt chatbots ``\textit{are very interesting and a big part of the future but can also be very dangerous because at the same time it is still programmed and could be set to give out any kind of information whether it be misinformation or not}.'' Some participants indicated they felt ``\textit{indifferent with no preference for or against chatbots}'' [\textbf{P112FL30}] and some were reserved but acknowledged ``\textit{how crazy advanced AI is that we now have chatbots that are able to assist}'' [\textbf{37FM30}].

Participants in the implicit ChatGPT group were unaware that the videos they viewed had text created by a language model and were much more likely to respond that they were unsure of their opinion on chatbots and language models. As indicated in Table \ref{tab:reception}, many of the participants in this group based their opinion of chatbots on the reliability of the information produced, noting that ``\textit{believe that they are the future of improving the lives of everyone and improving productivity exponentially}'' [\textbf{P69ML30}]. Some of the participants said that chatbots ``\textit{feel cheap and show a lack of effort from the creator; They are also impersonal}'' [\textbf{P80FM40}], commenting that they ``\textit{think chatbots are overblown and hyped up way too much; They still have humans shadow working on them constantly to make sure they preform as expected, feed them data and correct them}'' [\textbf{P95FL40}]. Participants also had a hesitant opinion of chatbots as ``\textit{they can be a great source of information, or a dangerous source of misinformation}'' [\textbf{P127FL20}]. 

\section{Discussion} \label{sec:discussion}
\subsection{Implications}
Aware of the obvious threat of causing physical harm by synthesizing speculative health responses, ChatGPT was designed to opine with a default recommendation for consulting a health professionals before attempting to consume the ``at-home aboriton'' herbs. We did not attempt to circumvent these settings, but it becomes increasingly easier to do so, as evidence show exploits of ChatGPT prompts to ``do anything now'' could anyhow produce intentionally manipulative and harmful responses \cite{Oremus}. The problems don't stop here, as our results suggest that even the default, socially responsible answers from ChatGPT are perceived as incomplete, lacking credibility, dangerous, unsafe, and scientifically unproven.  

Generating a plausibly sound but incorrect information even if trained on factual data is an acknowledged and open issue for LLMs and chatbots \cite{Brown} and hopefully the accuracy of the generated information will improve with time. But another open issue is the users' confidence in generative health and abortion information. Despite the good performances chatbots show on medical licensing tests or when writing basic diagnoses \cite{Kung, Gilson}, our findings show that a considerable number of users feel the chatbot simply pulls information of the Internet about ``at-home'' abortion herbs and already know not to trust what ChatGPT says. True, one could argue that we did not exposed our participants to directly interact with ChaptGPT but they nonetheless evaluated the accuracy of the chatbot's responses and noticed that the chatbot avoids doubling down on the harms and attempts to respond in balanced if somewhat grammatically incorrect manner. 

It is reassuring that a non-negligible number of entire sample pointed out they would fact-check and do their own search about what is a save abortion practice, now that the legal rights have been revoked in the United States. Some of our participants noted they would comment on the ChatGPT's response elements they believed are truthful and the elements they believed are misleading to help other TikTok users. This is a commendable engagement strategy and reinforces the previous evidence suggesting that TikTok facilitates social support exchange on thorny societal and health issues \cite{Barta, Southerton}. 

While the precarious labeling of our posts as misinformation by TikTok might be seen as overly intrusive, it might be a result of an increased algorithmic moderation in response to the criticisms that the platform pumps dangerous abortion content to young users \cite{newsguard}. Like before, the misinformation labels mattered little and the participants largely ignored it \cite{TikTok}, except in the case where ChatGPT explicitly produces a rap about the abortifacient herb pennyroyal. We are on the opinion that the perceived oddity of the lyrical response, coupled with the smartphone interface might have be the main confounding effect, but we are nonetheless content that the participants distanced themselves from the videos.

How TikTok and other social media platforms will handle generative health information remains to be seen, especially when it comes to moderating health and abortion misinformation. Currently, all mainstream platforms either use ``context labels'' to human-generated abortion content (YouTube and TikTok \cite{YouTube, Keenan}), promote authoritative abortion information (Twitter \cite{Kern}) or simply block questionable abortion treatment advertisements (Meta \cite{Meta}). Using chatbots and language models in strict medical circumstances is certainly lower on magnitude in generated content and likely supervised by health professionals that have to sign off the treatments or diagnoses, but on a social media scale this is impractical. Abusing chatbots for synthesized health misinformation, coupled with the capabilities for generative propaganda and rumors as in the case with GPT-4chan is a real threat \cite{Murphy}. In our opinion, this threat won't be entirely mitigated by users' early distrust in chatbots signaled in our study and the ``neural'' fake have the potential to spur a social media infodemic comparable to the chaos of COVID-19 misinformation \cite{TrollHunter}.

\subsection{Ethical Considerations}
While we debriefed our participants about the dangers of ``at-home'' abortion remedies, we acknowledge that there might be a potential risk of repeated exposure to abortion misinformation, i.e. an ``implied truth effect'' \cite{Pennycook-Rand-Psych}, as each participant saw four of the videos. To mitigate this risk we explicitly pointed to the debunked information for the related ``at-home'' remedies and their associated harms. Another risk that stems from our study is the possibility that participants can attempt or promote prompting ChatGPT and other chatbots about many different types of health information and untested medical practices. We warned that the study does not promote nor advocate for using generative recommendations of treatments or diagnoses without a consulting a healthcare professional. 

We are aware that studies with generative health and abortion (mis)information might risk oversimplification or misinterpretation of the findings, therefore we deliberately avoided providing definitive numbers beyond the participants' self-reported age, gender, and political leanings in our reporting. Abortion misinformation could have dangerous consequences and is regularly used in polarized political discourses revolving around the legality and availability of safe abortion treatments. For example, \#OpJane is the latest online operation launched against the state of Texas for enacting the anti-abortion Bill 8 that allows ``abortion bounty'' for anyone who will investigate and report abortion \cite{Goforth}. Because the operation calls for ``fighting misinformation with enough plausible and difficult to disprove misinformation'' to make any data these bounty hunters gather as useless \cite{OperationJane-Anon}, we exercise caution in exploiting  ChatGPT or other chatbots with ``do anything now'' jailbreak prompts to facilitate this call.

\subsection{Limitations}
Our research was limited in its scope to U.S. TikTok media users and the state of generative capabilities of ChatGPT regarding at-home abortion remedies at the time of the study. In as much as we attempted to use as close as possible generic abortion information prompts, other prompts to ChatGPT or another chatbot and LLM might produce different responses than the ones in our study.  A limitation also comes from the sampling method and the use of  Prolific as a participant recruit provider, as other users and other samples might provide results that differ from the once we obtained. We did not measure the efficacy of users' assessments and engagement strategies for human-generated abortion misinformation content, nor did we ask how users dealt with other abortion misinformation on other social media platforms. Short-form videos are a relatively new way of persuasive communication appealing to younger users, and generative use of images for creating ``neural memes'' or abortion misinformation ``deep fakes'' might provoke different responses for a wider population of users \cite{Struppek}. Therefore, we are careful to avoid any predictive use of our findings due to the malleability of the ``neural'' multimedia misinformation synthetization.

\section{Conclusion} \label{sec:conclusion}
Self-induced terminations of unwanted pregnancies in a post-\textit{Roe v Wade} America will undoubtedly drive many interested users to try chatbots for an advise regarding ``at-home'' abortion. Whether the generative responses will propagate on social media platforms verbatim or will be modified by humans to fit particular narratives is yet to be seen, but the barriers are already low for any questionably content on abortion to reach wide audiences. Our findings are reassuring that users on TikTok are able to distance themselves from generative responses and assume more of a ''better-safe-than-sorry'' position in engaging with them. The misinformation moderation strategies employed by the platforms and the availability of scientifically debunked abortion claims certainly affect this position and we hope that our study adds in the effort to prevent harmful outcomes of abortion misinformation.

\bibliographystyle{plain}
\bibliography{\jobname}

\newpage
\section*{Appendix}

\section*{Study Questionnaire} \label{sec:survey}
\subsection*{Engagement Strategies}
\begin{enumerate}
\itemsep 1em
        \item How do you suspect or know that a certain TikTok post is a misinformation? Please elaborate. \textbf{[Open Ended]}
        \item What is your strategy for dealing with misinformation posts on TikTok? Please elaborate. \textbf{[Open Ended]}             
\end{enumerate}

\subsection*{Language Modeling Exposure}
\begin{enumerate}
\itemsep 1em
    \item Have you encountered other videos created with Chatbots or Language Models? Please elaborate. \textbf{[Open Ended]}
    \item Have you used or experimented with Chatbots or Language Models? Please elaborate. \textbf{[Open Ended]}
    \item What is your general opinion of Chatbots or Language Models? Please elaborate. \textbf{[Open Ended]}
\end{enumerate}

\section*{ChatGPT Prompts and Responses} \label{sec:ChatGPT}

\subsection{Video 1: Abortifacient herbs} \label{sec:video1}

\begin{table}[!h]
\renewcommand{\arraystretch}{1.3}
\footnotesize
\label{tab:video1-response}
\centering
\aboverulesep=0ex 
   \belowrulesep=0ex 
\begin{tabularx}{1\linewidth}{|X|}
\Xhline{3\arrayrulewidth}
\toprule
 \textbf{Prompt:} Abortifacient herbs \\\Xhline{3\arrayrulewidth}
\midrule
\vspace{0.2em}
\textbf{Response:} Abortifacient herbs are plants that have been traditionally used to induce abortion or termination of pregnancy. Examples include pennyroyal, rue, and tansy. However, it is important to note that using these herbs can be dangerous and can cause serious health complications. It is also illegal in many countries to use these herbs for this purpose. It is strongly recommended to consult with a medical professional before using any abortifacient herbs. \\\Xhline{3\arrayrulewidth}
\end{tabularx}
\end{table}


\subsection{Video 2: Tell me about herbs for abortion} \label{sec:video2}

\begin{table}[!h]
\renewcommand{\arraystretch}{1.3}
\footnotesize
\label{tab:video2-response}
\centering
\aboverulesep=0ex 
   \belowrulesep=0ex 
\begin{tabularx}{1\linewidth}{|X|}
\Xhline{3\arrayrulewidth}
\toprule
 \textbf{Prompt:} Tell me about herbs for abortion \\\Xhline{3\arrayrulewidth}
\midrule
\vspace{0.2em}
\textbf{Response:} It is important to note that inducing an abortion through the use of herbs can be dangerous and is not recommended by medical professionals. If you are pregnant and considering an abortion, it is important to speak with a healthcare provider about safe and legal options. \\ 

Herbs that have been traditionally used for inducing abortion including pennyroyal, tansy, and goldenseal. However, these herbs can be toxic and have been linked to serious health complications, including death. The use of these herbs should be avoided. \\

It is important to remember that self-induced abortion can be dangerous and is illegal in some areas. It is always best to consult a healthcare professional for guidance and information about safe and legal options. 
\\\Xhline{3\arrayrulewidth}
\end{tabularx}
\end{table}


\subsection{Video 3: Tell me all of the facts about pennyroyal} \label{sec:video3}

\begin{table}[!h]
\renewcommand{\arraystretch}{1.3}
\footnotesize
\label{tab:video3-response}
\centering
\aboverulesep=0ex 
   \belowrulesep=0ex 
\begin{tabularx}{1\linewidth}{|X|}
\Xhline{3\arrayrulewidth}
\toprule
 \textbf{Prompt:} Tell me all of the facts about pennyroyal  \\\Xhline{3\arrayrulewidth}
\midrule
\vspace{0.2em}
\textbf{Response:} Pennyroyal is a perennial herb in the mint family, scientifically known as Mentha pulegium. It is native to Europe and Asia and has been used for centuries in traditional medicine for a variety of purposes, including as a treatment for respiratory and digestive issues and as a menstrual aid. It is also sometimes used as a culinary herb and insect repellent. However, it is important to note that pennyroyal is toxic in large doses and should not be consumed in large quantities or used as a self-treatment without consulting a healthcare professional. Some of the toxic effects of pennyroyal include liver and kidney damage, seizures, and even death. \\\Xhline{3\arrayrulewidth}
\end{tabularx}
\end{table}

\subsection{Video 4: Write a rap about pennyroyal} \label{sec:video4}

\begin{table}[!h]
\renewcommand{\arraystretch}{1.3}
\footnotesize
\label{tab:video4-response}
\centering
\aboverulesep=0ex 
   \belowrulesep=0ex 
\begin{tabularx}{1\linewidth}{|X|}
\Xhline{3\arrayrulewidth}
\toprule
 \textbf{Prompt:} Write a rap about pennyroyal  \\\Xhline{3\arrayrulewidth}
\midrule
\vspace{0.2em}
\textbf{Response:} Listen up y’all, I got a herb to tell
        Pennyroyal, it’s a gem, it’ll make you well
        Used for centuries, it’s a natural cure
        For a variety of ailments, it’s pure \\

        Menthol scent, it’s minty and fresh
        Used for menstrual cramps, it’ll give you some rest
        Tea or oil, take it as you please
        Pennyroyal, it’ll ease your disease \\

        But be careful, don’t overdo it
        Too much can be toxic, don’t you see it
        Moderation is key, don’t abuse
        Pennyroyal, it’ll give you the blues \\

        So if you’re feeling down, give it a try
        Pennyroyal, it’ll make you fly
        Natural medicine at its best
        Pennyroyal it’ll pass the test
\\\Xhline{3\arrayrulewidth}
\end{tabularx}
\end{table}

\newpage
\end{document}